\documentclass[conference]{IEEEtran}

\usepackage{cite}
\usepackage{amsmath,amssymb,amsfonts}
\usepackage{algorithmic}
\usepackage{graphicx}
\usepackage{textcomp}
\usepackage{xcolor}
\usepackage[hyphens]{url}

\usepackage{ulem}
\usepackage{soul}  
\usepackage{multirow}
\usepackage{enumitem}
\usepackage{lipsum}
\usepackage{array}
\usepackage{subcaption}
\setlipsum{
  par-before = \begingroup\color{gray},
  par-after = \endgroup
}

\usepackage[bookmarks=true,breaklinks=true,letterpaper=true,colorlinks,citecolor=blue,linkcolor=blue,urlcolor=blue]{hyperref}

\newcommand{\insertFigure}[5]{
    \begin{figure}[t]
      \centering
      \includegraphics[width=#3\linewidth]{figure/#1}
      \vspace{#4}
      \caption{\small #2}
      \label{fig:#1}
      \vspace{#5}
    \end{figure}
}

\newcommand{\insertSubFigure}[5]{
    \begin{figure}[t!]
        \centering
        \begin{subfigure}[t]{0.43\linewidth}
            \centering
            \vspace{-1.2em}
            \includegraphics[width=\textwidth]{figure/#1}
            \vspace{-2.3em}
            \caption{#2}
        \end{subfigure}
        \begin{subfigure}[t]{0.43\linewidth}
            \centering
            \vspace{-2.6em}
            \includegraphics[width=\textwidth]{figure/#3}
            \\
            \vspace{-1.2em}
            \caption{#4}
        \end{subfigure}
        \caption{#5}
        \label{fig:evalschemas}
        \vspace{-1em}
    \end{figure}
}
\newcommand{\msccl}{MSCCL}
\newcommand{\mscclang}{MSCCLang}
\newcommand{\taccl}{TACCL}
\newcommand{\tacos}{TACOS}

\newcommand{\mscclir}{MSCCL-IR}

\newcommand{\chakra}{Chakra ET}
\newcommand{\astrasim}{ASTRA-sim}

\newcommand{\allreduce}{All-Reduce}
\newcommand{\allgather}{All-Gather}

\newcommand{\ring}{Ring}
\newcommand{\doubleBinary}{Double Binary Tree}

\newcommand{\csend}{\textsc{COMM\_SEND}}
\newcommand{\crecv}{\textsc{COMM\_RECV}}
\newcommand{\ccomp}{\textsc{COMP}}

\def\BibTeX{{\rm B\kern-.05em{\sc i\kern-.025em b}\kern-.08em
    T\kern-.1667em\lower.7ex\hbox{E}\kern-.125emX}}
\begin{document}

\title{Towards a Standardized Representation\\for Deep Learning Collective Algorithms}

\author{
    \IEEEauthorblockN{
        Jinsun Yoo\IEEEauthorrefmark{1}, William Won\IEEEauthorrefmark{1}, Meghan Cowan\IEEEauthorrefmark{2}, Nan Jiang\IEEEauthorrefmark{2},\\
        Benjamin Klenk\IEEEauthorrefmark{2}, Srinivas Sridharan\IEEEauthorrefmark{2}, and Tushar Krishna\IEEEauthorrefmark{1}
    }
    \IEEEauthorblockA{
        \IEEEauthorrefmark{1}Georgia Institute of Technology~~~~\IEEEauthorrefmark{2}NVIDIA
    }
    \IEEEauthorblockA{
        \IEEEauthorrefmark{1}\{jinsun, william.won\}@gatech.edu, tushar@ece.gatech.edu~~~
        \IEEEauthorrefmark{2}\{mcowan, tedj, bklenk, srisridharan\}@nvidia.com
    }
}

\maketitle

\begin{abstract}

The explosion of machine learning model size has led to its execution on distributed clusters at a very large scale. Many works have tried to optimize the process of producing collective algorithms and running collective communications, which act as a bottleneck to distributed machine learning. However, different works use their own collective algorithm representation, pushing away from co-optimizing collective communication and the rest of the workload. The lack of a standardized collective algorithm representation has also hindered interoperability between collective algorithm producers and consumers. Additionally, tool-specific conversions and modifications have to be made for each pair of tools producing and consuming collective algorithms which adds to engineering efforts. 

In this position paper, we propose a standardized workflow leveraging a common collective algorithm representation. Upstream producers and downstream consumers converge to a common representation format based on Chakra Execution Trace, a commonly used graph based representation of distributed machine learning workloads. Such a common representation enables us to view collective communications at the same level as workload operations and decouple producer and consumer tools, enhance interoperability, and relieve the user from the burden of having to focus on downstream implementations. We provide a proof-of-concept of this standardized workflow by simulating collective algorithms generated by the \mscclang\ domain-specific language through the \astrasim\ distributed machine learning simulator using various network configurations.

\end{abstract}

\begin{IEEEkeywords}
distributed ML systems, collective communication, collective algorithm, simulation
\end{IEEEkeywords}

\section{Introduction}\label{sec:introduction}

Recent trends in enormous machine learning (ML) models, such as recommendation models~\cite{naumov2019dlrm} or Large Language Models (LLMs)~\cite{brown2020gpt3}, have made it impractical to execute them on a single Neural Processing Unit (NPU, such as GPU, TPU, or custom ASIC). Consequently, ML execution has evolved to \textit{distribute} the job across multiple NPUs. Within distributed ML, each of the participating NPUs completes a portion of the overall compute task. They periodically transfer and synchronize their intermediate compute results (e.g., weight or input gradients) in accordance with a predefined schedule~\cite{tang2023survey}. Such traffic patterns are \textit{collective} in nature. These \textit{collective communications}, such as \allreduce\ or \allgather, have been the key building blocks of distributed ML platforms~\cite{klenk2020allreduce}.

The communication of intermediate data has become a bottleneck to the overall distributed ML execution, and recent studies have tried to optimize this~\cite{rashidi2021ace, sapio2021bottleneck, won2024libra}. Within the scope of collectives, Collective Communication Libraries (CCLs) such as NVIDIA's NCCL~\cite{nvidianccl} provide implementations of several predefined \textit{collective algorithms} (e.g., \ring\ and \doubleBinary). Several works have tried to further optimize collective algorithms, by designing complex algorithms~\cite{thakur2005mpich}, automatically \textit{synthesizing} topology-aware algorithms~\cite{taccl, won2024tacos}, or allowing users to easily define and test their own~\cite{cowan2023mscclang}.

\insertFigure{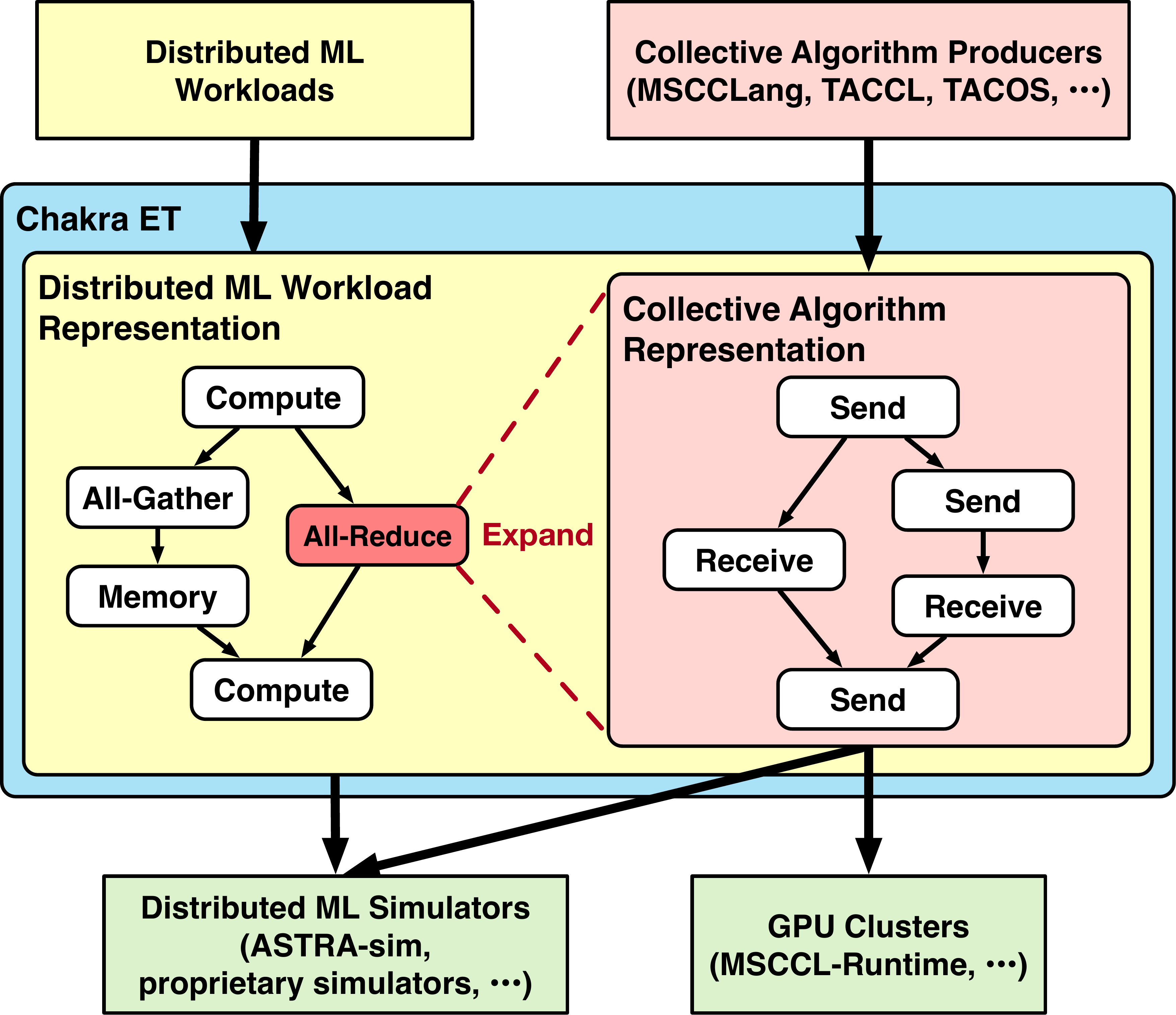}{
The proposed standardized workflow using Chakra Execution Trace (\chakra) as a common representation for both distributed ML workload and collective algorithm. Downstream tools receive both workload and collective algorithms represented using a common \chakra\ format. Sample \chakra s of the workload (left) and that of the algorithm of a single collective (right) is also shown.
}{1}{-1em}{-1.5em}

\textbf{Unfortunately, these collective optimizations have been done separately from distributed ML workload optimization}, due to a lack of common representation. Specifically, distributed ML workload information does not contain details on how each collective operation is implemented, and rely on tools such as simulators or cluster runtimes to fill in the blank on their own. As such, these tools have their own implementation of collective algorithms. \textbf{To the best of our knowledge, there is no common format that represents both the distributed ML workloads and collective algorithms}. Because of this, collective communications and other operators within a workload (such as compute or memory access) are produced, consumed, and optimized separately. 

To bridge this gap, \textbf{we propose a common representation format for both workload and collective algorithm across different tools}. We draw attention to the Chakra Execution Trace (\chakra) format, a standard graph-based representation of execution traces for distributed ML workloads~\cite{sridharan2023chakra}. In this work, we extend \chakra\ to also encode collective algorithms. \autoref{fig:hierarchy} summarizes the idea, where (i)~both the distributed ML workload and collective algorithm are represented in a common \chakra\ format and (ii)~downstream tools can directly ingest \textit{a single} \chakra\ representation for \textit{both} simulation or execution. We develop and opensource a proof-of-concept workflow where users write custom collective algorithms using the \mscclang\ Domain Specific Language\cite{cowan2023mscclang}, represent them in \chakra\ format, and compare the algorithms using the \astrasim\ distributed ML system simulator\cite{won2023astrasim2}.\footnote{\url{https://github.com/astra-sim/collectiveapi}} Similar workflows can be easily built leveraging this standard.

There are three key benefits to employing \chakra\ as a common representation. First, we envision \textbf{co-optimizing collective communication with other operations in the workload via leveraging a streamlined workflow} encapsulating both the workload and collective algorithms. Second, \textbf{interoperability across different tools} allows users to leverage a single representation to simulate a collective algorithm on simulators like \astrasim~\cite{won2023astrasim2} and validate the \textit{same} algorithm on real systems like \msccl-Runtime~\cite{cowan2023mscclang}. Finally, the decoupling between tools and the representation helps \textbf{reduce the effort to implement collective algorithms for each downstream tool}. For example, \astrasim\ could ingest the common representation instead of users having to implement the collective algorithm within its codebase, which requires \textit{simulator-specific} knowledge.

To summarize, our contributions are as follows:
\begin{itemize}[leftmargin=*]
    \item We motivate and propose a standard workflow that uses a common collective algorithm representation to bridge distributed ML workload information, upstream collective algorithm producers, and downstream tools.
    \item As proof-of-concept, we provide a case study showing interoperability with \mscclang\ and the \astrasim\ simulator.
\end{itemize}

\section{Background}\label{sec:background}

\subsection{Chakra Execution Trace}

\chakra~\cite{sridharan2023chakra} aims to provide a standard graph-based representation to capture the trace of distributed ML workload execution. It represents a distributed workload using a directed acyclic graph whose vertices denote ML operations and edges indicate inter-operation dependencies. \chakra\ can be fed into and consumed by distinct downstream tools, such as distributed ML simulators or benchmarking tools. These downstream tools leverage \chakra\ by traversing through the graph and issuing the operations whose dependencies are resolved and are ready to be dispatched. \chakra\ graphs can be collected in multiple ways, for instance, through profiling actual PyTorch executions or via synthetic generations.

Note that a collective communication node in \chakra\ merely indicates that a collective communication has taken place. However, it does not encode the actual collective algorithm. In other words, \chakra\ itself is oblivious to exactly how the messages are orchestrated and transferred. Consequently, the downstream tools exploit their internal, tool-specific implementations of collective algorithms.

\subsection{Upstream Collective Algorithm Producers}\label{subsec:upstream}

Given the significance of collective communication in distributed ML, numerous studies have focused on optimizing collective algorithms. Primarily, they have pursued two main threads: (i)~developing domain-specific languages (DSLs) to enable users to define their own collective algorithms and (ii)~implementing synthesizers to autonomously generate them.

One notable example is \mscclang~\cite{cowan2023mscclang}, which introduces a Python-based DSL for collective algorithms. This enables users to easily construct NCCL-based collective algorithms. \mscclang\ compiles these algorithms into an XML-based representation (\mscclir), which is then executed on real clusters via the NCCL-based \msccl-Runtime. Meanwhile, synthesizers like \taccl~\cite{taccl} and \tacos~\cite{won2024tacos} generate collective algorithms tailored to the network topology. \taccl\ employs Integer Linear Programming (ILP) to identify optimal collective algorithms, while \tacos\ utilizes a Time-expanded Network (TEN) approach.

\subsection{Downstream Distributed Machine Learning Tools}\label{subsec:downstream}

Downstream tools receive distributed ML workloads or collective algorithm representations to execute meaningful tasks. Collective communication runtimes and distributed ML simulators are notable instances of such downstream tools. For instance, the \msccl-Runtime orchestrates a collective communication by taking the collective algorithm in \mscclir\ format and executing it on real GPU clusters via an NCCL-based runtime. Conversely, \astrasim\ is a notable example of simulation infrastructure. Its Workload Layer can receive and simulate a distributed ML workload in \chakra\ format, while its System Layer implements collective algorithms to fill in the gaps of the workload \chakra. The Network Layer captures requests from these layers and simulates the network transfers over a network simulator of user choice.

\section{Collective Algorithm Representation}

\subsection{Motivation: Needs for Standardization}

Currently, upstream collective algorithm producers employ unique representations to describe their results. For example, \mscclang\ utilizes a NCCL-based, low-level XML interpretation, while \tacos\ relies on its own TEN representation.

This lack of standardization often leads downstream tools to rely on their \textit{unique internal implementation}. Consequently, the format and pipeline that they use to fetch collective information diverge from those used to inject workload information. As a result, users are constrained to optimizing either collective operations or other workload operations, not both.

Moreover, the absence of a standard format means that executing an upstream producer's algorithm with a specific downstream tool requires users to comprehend the internal details of both tools and implement the algorithm themselves. This task is not only highly prohibitive but also implies that it must be repeated for every pair of upstream and downstream tools. Consequently, upstream and downstream processes become disjointed and lack plug-and-play functionality.

\begin{scriptsize}
\begin{table}[t!]
\centering
\caption{A list of \chakra\ node types used to represent collective algorithms and their description.}
\vspace{-0.5em}
\label{table:chakra_nodes}

\begin{tabular}{|c|l|}
    \hline
    \textbf{\chakra\ Node Type} & \textbf{Description} \\ \hline
    \csend & (Pt-to-pt) Message send to a destination. \\ \hline
    \crecv & (Pt-to-pt) Wait for a message from a source. \\ \hline
    \ccomp & \begin{tabular}[c]{@{}l@{}}Run a compute task (e.g., Reduction)\end{tabular} \\ \hline
\end{tabular}

\vspace{-2em}
\end{table}
\end{scriptsize}

\subsection{Solution: Using Chakra Execution Trace}

We standardize the collective algorithm representation by utilizing the \chakra\ format that is already employed for distributed ML workloads. It readily offers mechanisms to capture point-to-point message transfers between arbitrary NPUs as well as compute operations necessary for collectives.

By representing both distributed ML workloads and collective algorithms in \chakra, we cleanly resolve the issue of separation among workloads, upstream, and downstream tools. \autoref{fig:hierarchy} depicts the proposed standardized workflow. Both distributed ML workloads and collective algorithms are in \chakra\ format and passed on to the downstream. The downstream tool then traverses the workload ET and executes operations as their dependencies are resolved. During the process, since collective communication nodes lack the exact mechanism to execute collectives, the downstream tools must decide the algorithm. Previously, they selected a collective algorithm from a range of native implementations or custom algorithms tailored to their specifications. However, with the capability to easily receive any collective algorithm in \chakra, the downstream tools can expand the collective communication node with the provided algorithm. Users can rapidly test new collective algorithms by simply switching out the \chakra\ files without creating new tool-specific implementations.

By representing collective algorithms with the same format as the workload, we elevate the send and receive messages of a collective algorithm to the same level as other operators in the workload. This opens up opportunities for co-optimizing collective communication and other operators such as compute. For example, it is much simpler to test a scheduling feature that reorders a compute node and a send node (assuming the reorder respects inter-node dependencies) as the compute and send node now use a common \chakra\ format. 

Note that real-world systems such as \msccl-Runtime do not take workload information as input. Even in this case, having the \chakra\ as a common collective algorithm representation helps bridge the gap with upstream producers. For example, a user may want to simulate multiple prospective algorithms using \astrasim, then validate the best-performing candidates by actually using the \msccl-Runtime. The user in this workflow can reuse the same \chakra\ format across both tools without modification. This is achieved by the fact that the common representation abstracts away the details of the downstream tools.

\insertSubFigure{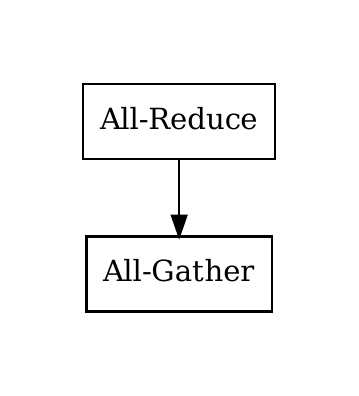}{Distributed ML Workload Representation}{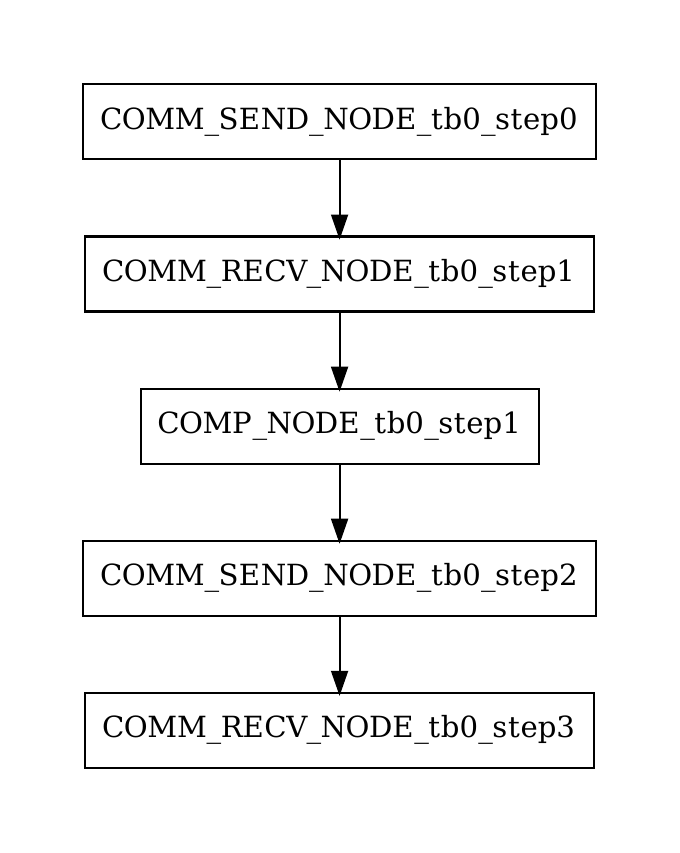}{Part of the Collective Algorithm Representation}{Snippets of the \chakra\ used to represent the workload and \ring\ collective algorithm used in the evaluation.}

\subsection{Collective Algorithm in \chakra}

\autoref{table:chakra_nodes} lists the types of \chakra\ operator nodes that are used to represent collective algorithms and their definition. Leveraging these nodes, \chakra\ can represent point-to-point network transfer between two NPUs and compute operations. Arbitrary collective algorithms can then be described as a combination of network message transfers and reductions.

To meet the standardization requirement, upstream tools need to convert resulting collective algorithms from their default representation to the common \chakra\ format when producing collective algorithms. We highlight that implementing this conversion is a one-time task such that, once developed, can be reused across multiple downstream frameworks.

\section{Methodology}\label{sec:methodology}

As a case study and proof-of-concept, we construct an end-to-end workflow using \mscclang\ and \astrasim. Here, we describe the extensions made to the two tools for the case study. We expect similar extensions would apply to other tools.

\subsection{Representing \mscclang\ Output in \chakra}

We developed a converter that bridges the \mscclir\ format into \chakra. The converter creates a vertex (i.e., \chakra\ node) for each operation in the \mscclir\ format. The converter then creates edges by extracting inter-operator dependency information encoded in the \mscclir\ format. To showcase the update, we described a 1D Ring algorithm of \allreduce\ using \mscclang\ and compiled the result in the standard \chakra\ format. \autoref{fig:evalschemas}(b) shows part of the \chakra-based 1D Ring algorithm representation.

\subsection{Updating \astrasim\ to Run Algorithms in \chakra}

\insertFigure{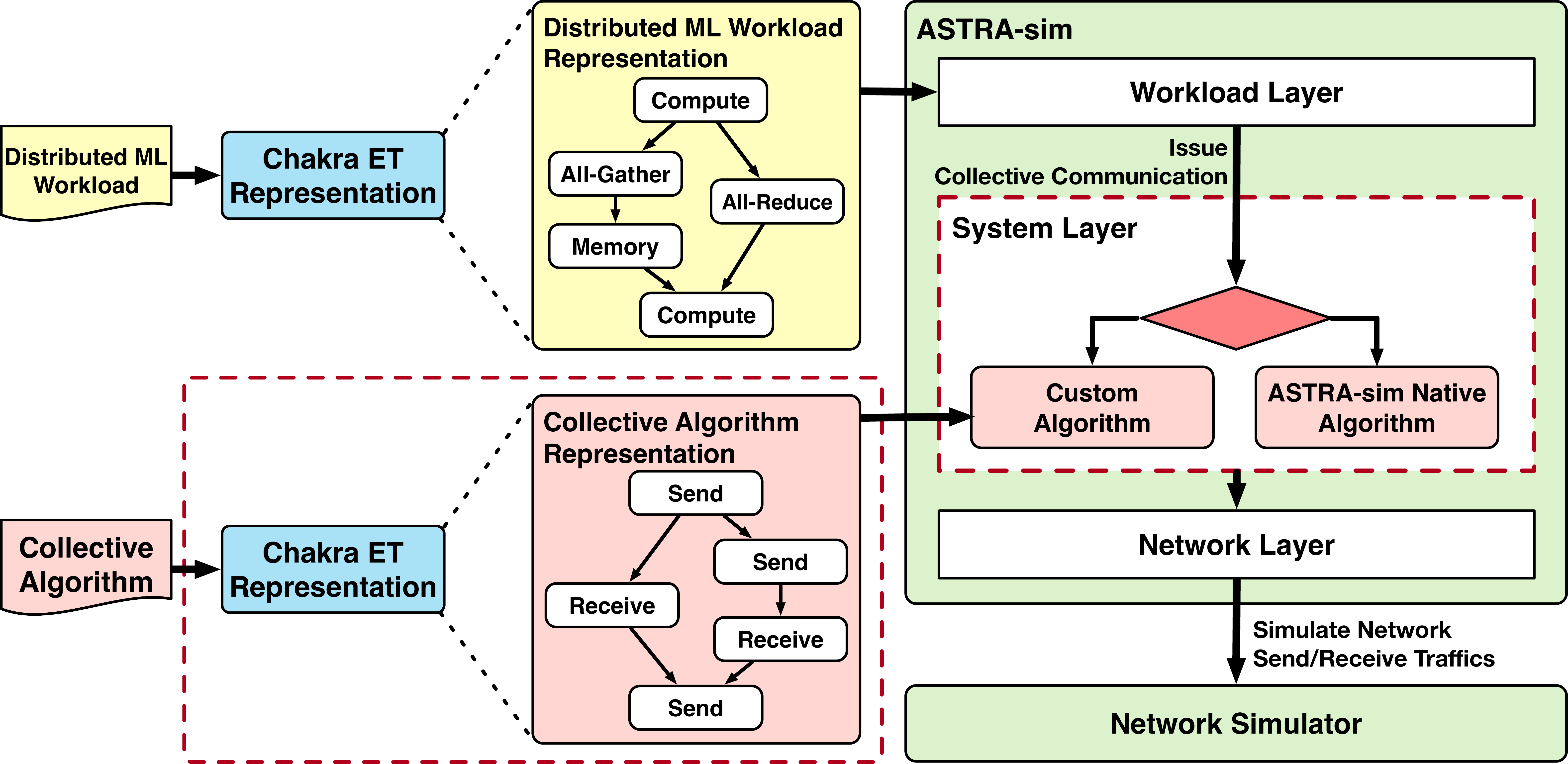}{
Components of \astrasim\ involved in the case study. We extended \astrasim\ to inject collective algorithms represented in \chakra. The extension is marked with dashed squares.
}{1}{-1.5em}{-0.5em}

\autoref{fig:schema} shows the workflow of \astrasim\ and our extensions. \astrasim\ includes its implementation of collective algorithms found in NCCL, such as \ring, out of the box. At each run, the user chooses which algorithm to use for each collective. Whenever a collective communication node is issued, \astrasim\ runs the corresponding algorithm code. 

We extend \astrasim\ by adding an input parameter that accepts the \chakra\ representation of the collective algorithm. The simulator parses the provided \chakra\ and simulates the point-to-point send and receive commands following the dependencies, rather than using its own implementation. This process is depicted in dashed boxes. Since \astrasim\ readily supports the execution of ML workloads in \chakra, reusing the components in the workload layer has made it easy to extend for the common collective representations as well.

\section{Evaluation}\label{sec:evaluation}

We showcase our case study of bridging \mscclang\ and \astrasim\ as described in \autoref{sec:methodology}. We use the workload depicted in \autoref{fig:evalschemas}(a) and use \mscclang\ to generate a 1D \ring\ algorithm for both collectives. \autoref{fig:evalschemas} shows a snippet of the \chakra\ represented as a graph. Note how we can represent both the workload and the collective algorithm using the same \chakra\ format. The resulting \chakra\ is then provided to the \astrasim\ simulator to be tested across different topologies with varying physical connectivity.

\insertFigure{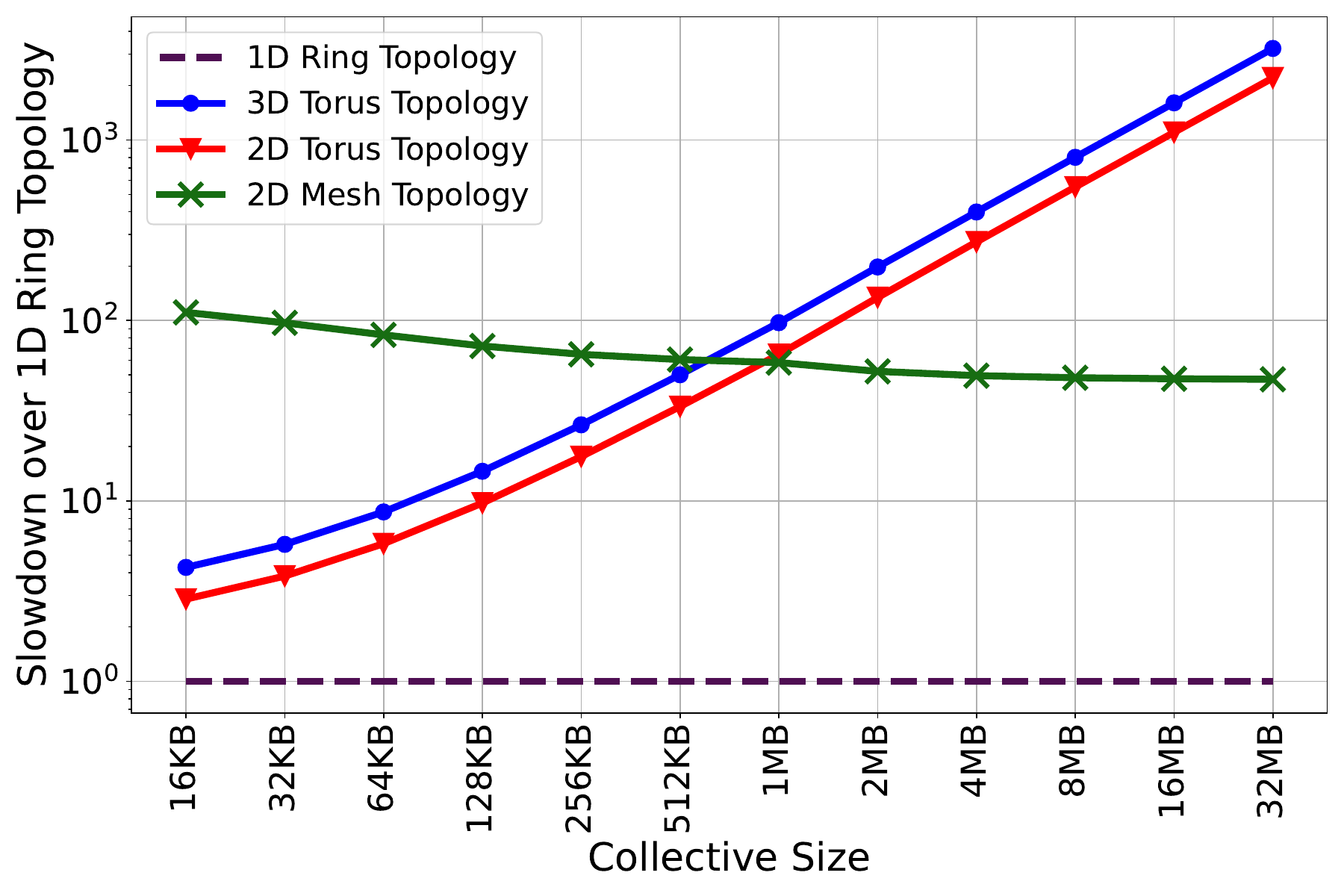}{
 The collective duration for a 1D Ring algorithm across different topologies of 64 NPUs.
}{0.75}{-1em}{-1.5em}

We use the Analytical network simulator to model the message transfer. The topologies consist of 64 NPUs with varying connectivity. We observe the workload duration as we differ the collective size. \autoref{fig:eval} shows the slowdown of the different topologies compared to a 1D \ring. Such an experiment was made possible thanks to the streamlined workflow via standardization of the collective algorithm representation. It is natural that other topologies show slowdowns as we use a simple \ring\ algorithm for both \allreduce\ and \allgather. While this evaluation showcases a simplistic workload, it is possible to expand into complex workloads and collective algorithms. We leave this to future work.

\section{Conclusion and Future Work}\label{sec:conclusion}

This paper proposes a standardized common representation for collective algorithms. We reuse the \chakra\ format, which already captures distributed ML workload traces, as the collective representation. Representing both workloads and collective algorithms with the same format will allow us to explore the co-optimization of workload operators and collective communication operators. We showcase such a common representation with a case study using collective algorithms produced by \mscclang\ on the \astrasim\ simulator.

Note that our proposition is focused on standardizing representations. Tasks such as design space exploration for collective algorithms or parallelization strategy are left to other tools. Our work simply provides a standard representation which opens rooms for co-optimization and interoperability.

Our work opens up several future research directions. One important future work is to leverage the proposed standard representation to explore collective optimizations in the context of an actual workload. We anticipate that a common representation across workloads and collective algorithms will allow us to study the co-optimization, such as overlapping, of compute and communication operations. Another potential direction would be to enable other tools to produce and consume \chakra\ format, further expanding the ecosystem and scope of our proposed workflow.


\bibliographystyle{IEEEtran}
\bibliography{reference/reference.bib}

\begin{thebibliography}{10}
\providecommand{\url}[1]{#1}
\csname url@samestyle\endcsname
\providecommand{\newblock}{\relax}
\providecommand{\bibinfo}[2]{#2}
\providecommand{\BIBentrySTDinterwordspacing}{\spaceskip=0pt\relax}
\providecommand{\BIBentryALTinterwordstretchfactor}{4}
\providecommand{\BIBentryALTinterwordspacing}{\spaceskip=\fontdimen2\font plus
\BIBentryALTinterwordstretchfactor\fontdimen3\font minus \fontdimen4\font\relax}
\providecommand{\BIBforeignlanguage}[2]{{%
\expandafter\ifx\csname l@#1\endcsname\relax
\typeout{** WARNING: IEEEtran.bst: No hyphenation pattern has been}%
\typeout{** loaded for the language `#1'. Using the pattern for}%
\typeout{** the default language instead.}%
\else
\language=\csname l@#1\endcsname
\fi
#2}}
\providecommand{\BIBdecl}{\relax}
\BIBdecl

\bibitem{naumov2019dlrm}
M.~Naumov \emph{et~al.}, ``{Deep Learning Recommendation Model for Personalization and Recommendation Systems},'' in \emph{arXiv:1906.00091}, 2019.

\bibitem{brown2020gpt3}
T.~Brown \emph{et~al.}, ``{Language Models are Few-Shot Learners},'' in \emph{NeurlIPS '20}, 2020, pp. 1877--1901.

\bibitem{tang2023survey}
Z.~Tang \emph{et~al.}, ``{Communication-Efficient Distributed Deep Learning: A Comprehensive Survey},'' in \emph{arXiv:2003.06307}, 2023.

\bibitem{klenk2020allreduce}
B.~Klenk \emph{et~al.}, ``{An in-network architecture for accelerating shared-memory multiprocessor collectives},'' in \emph{ISCA '20}, 2020, p. 996–1009.

\bibitem{rashidi2021ace}
S.~Rashidi \emph{et~al.}, ``{Enabling Compute-Communication Overlap in Distributed Deep Learning Training Platforms},'' in \emph{ISCA}, 2021, pp. 540--553.

\bibitem{sapio2021bottleneck}
A.~Sapio \emph{et~al.}, ``{Scaling Distributed Machine Learning with In-Network Aggregation},'' in \emph{NSDI '21}, 2021, pp. 785--808.

\bibitem{won2024libra}
W.~Won \emph{et~al.}, ``{LIBRA: Enabling Workload-aware Multi-dimensional Network Topology Optimization for Distributed Training of Large AI Models},'' in \emph{ISPASS '24}, 2024.

\bibitem{nvidianccl}
{NVIDIA}, ``{NVIDIA Collective Communications Library (NCCL)},'' \url{https://developer.nvidia.com/nccl}, 2024.

\bibitem{thakur2005mpich}
R.~Thakur \emph{et~al.}, ``{Optimization of Collective Communication Operations in MPICH},'' \emph{Int. J. High Perform. Comput. Appl.}, vol.~19, no.~1, p. 49–66, 2005.

\bibitem{taccl}
A.~Shah \emph{et~al.}, ``{TACCL: Guiding Collective Algorithm Synthesis using Communication Sketches},'' in \emph{NSDI '23}, 2023, pp. 593--612.

\bibitem{won2024tacos}
W.~Won \emph{et~al.}, ``{TACOS: Topology-Aware Collective Algorithm Synthesizer for Distributed Machine Learning},'' in \emph{arXiv:2304.05301}, 2024.

\bibitem{cowan2023mscclang}
M.~Cowan \emph{et~al.}, ``{MSCCLang: Microsoft Collective Communication Language},'' in \emph{ASPLOS '23}, 2023, pp. 502--514.

\bibitem{sridharan2023chakra}
S.~Sridharan \emph{et~al.}, ``{Chakra: Advancing Performance Benchmarking and Co-design using Standardized Execution Traces},'' in \emph{arXiv:2305.14516}, 2023.

\bibitem{won2023astrasim2}
W.~Won \emph{et~al.}, ``{ASTRA-sim2.0: Modeling Hierarchical Networks and Disaggregated Systems for Large-model Training at Scale},'' in \emph{ISPASS '23}, 2023, pp. 283--294.

\end{thebibliography}

\end{document}